# What Benefits Drive Membership in Medicare Advantage Plans?

Xiyue Liao, Ian Duncan, Jiarui Yu

**Abstract**
We seek to identify the most relevant benefits offered by Medicare Advantage Health Plans that drive membership and market share. As an example, we explore plans operating in a single county in New Jersey between 2018 and 2023. A dataset of benefits from publicly-available data sources was created and the variance inflation factor was applied to identify the correlation between the extracted features, to avoid multicollinearity and overparameterization problems. We categorized the variable Market Share and used it as a multinomial response variable with 3 categories: less than 0.3%; 0.3% to 1.5% and over 1.5%. Categories were chosen to achieve approximately uniform distribution of plans (47, 60 and 65 respectively). We built a multinomial Lasso model using 5-fold cross validation to tune the penalty parameter. Lasso forced some features to be dropped from the model, which reduces the risk of overfitting and increases the interpretability of the results. For each category, important variables are different. Certain brands drive market share, as do PPO plans and prescription drug coverage. Benefits, particularly ancillary benefits that are not part of CMS's required benefits appear to have little influence, while financial terms such as deductibles, copays and out-of-pocket limits are associated with higher market share. Finally, we evaluated the predictive accuracy of the Lasso model with the test set. The accuracy is 0.76.

## 1 Introduction

Medicare is the United States (US) federal health insurance program for people aged 65 or older, people under 65 with certain disabilities, and people of all ages with End-Stage Renal Disease. Medicare-eligible members receive their benefits through Original Medicare, a fee-for-service (FFS) system that includes most US providers, as well as Medicare-approved private insurers known as Medicare Advantage (MA) plans. Although MA plans have more limited provider networks increasing numbers of members have chosen to enroll in MA plans in recent years because they receive additional benefits, over and above those in traditional Medicare. Many plans are available to beneficiaries, and the range of additional benefits has also expanded. The plan quality (Star rating) system continues to foster competition between plans to improve quality in return for financial rewards. Private insurers compete by offering more attractive benefits, and members choose plans based on premiums, cost-sharing, benefits and Star ratings.

### 1.1 Medicare Advantage Plans

Medicare consists of four different parts, A (Hospital), B (Major Medical), C (Medicare Advantage or private insurer plans) and D (Drug benefits). MA plans include all benefits provided under Medicare Parts A and B. Additionally, most MA plans offer drug coverage (Part D) and may include an array of extra benefits such as dental, hearing, over-the-counter (OTC) medications, remote access technologies, transportation, and meal services. There are five types of MA Plans: Health Maintenance Organizations (HMOs), Preferred Provider Organizations (PPOs), Private Fee-For-Service (PFFS), Special Needs Plans (SNPs) and Employer Group Waiver Plans (EGWPs). For this study we focus on HMO/PPO plans. In 2023, nearly half (48%) of eligible Medicare beneficiaries were enrolled in MA plans, with enrollment figures doubling over the past fifteen years. The Congressional Budget Office (CBO) projects that the share of Medicare beneficiaries enrolled in MA plans will rise to 61% by 2032.

As enrollment in MA plans continues to grow, the number of plans available to beneficiaries has also increased. In 2023, 3,998 MA plans are available nationwide, and the average Medicare beneficiary can choose from plans offered by 9 insurers. About half of all Medicare beneficiaries (53%) (in 19% of counties) can choose from more than 40 MA plans.

### 1.1.1 *The Medicare Advantage Payment System*

MA plans are reimbursed in the form of a risk-adjusted premium per member. Understanding the financing of MA plans reveals a complex interplay of base rates, risk scores, premiums, quality bonuses and rebates. The payment to a plan is calculated as follows:

$$\text{Plan Revenue} = \text{Plan Base Rate} \times \text{Risk Score} + \text{Premium} + \text{Rebate (or Adjustment)}$$

Each year, plans submit bids to CMS that include proposed plan features and benefits and a proposed base rate representing the plan's estimated cost of covering a member in average health. CMS calculates a benchmark rate for the county in which the plan operates based on Medicare FFS experience. The base rate for revenue purposes is the lesser of the plan's bid and the benchmark rate. For MA plans that operate in more than one county, a weighted average benchmark is computed. Benchmarks are adjusted to reflect the quality of the plan as shown in table 1. Plans with 4 stars or more receive a 5% increase in their benchmark and new plans receive a 3.5% increase.

The second component of the plan's payment, the risk score reflects the health expenditure risk of the plan's enrollees. A member with average health is attributed to a risk score of 1.0. CMS uses versions of a risk adjustment model introduced by Pope et al. ([1, 2]). The third component is the premium, which is paid by the enrollee, set to 0 for plans whose bid is lower than the benchmark and as the difference between the bid and the benchmark for the other plans. Finally, plans that bid below the benchmark receive a rebate equal to a percentage of the difference between the benchmark and the bid. That is:

$$\text{Rebate} = (\text{Benchmark} - \text{Bid}) \times \text{Rebate Percentage}, \text{ where the rebate percentage is given in Table 1.}$$

The plan whose bid is greater than the benchmark does not receive a rebate. If its enrollees' average risk score is larger (smaller) than the overall average risk score the plan's payment is adjusted upwards (downwards) to maintain a fixed enrollee premium. This mechanism ensures that payments are aligned with both the plan's cost projections and market standards, adjusted for quality ratings and risk profiles of enrollees.

| Star Rating | CMS Benchmark | Rebate Percentage |
|---|---|---|
| 5.0 Star | 5% Bonus Rate | 70% |
| 4.5 Star | | 70% |
| 4.0 Star | | 65% |
| New | 3.5% Bonus Rate | 65% |
| 3.5 Star | 0% Bonus Rate | 65% |
| 3.0 Star or Less | | 50% |

**Table 1  Rebate percentages and benchmark increase**

*1.1.2 The Medicare Star Rating System*

An important component of both plan attraction and reimbursement is the plan's Star rating, which measures the performance of the plan with respect to certain quality measures. High-quality plans are allowed to enroll members throughout the year (rather than at an open enrollment time) and to receive additional funds that they are required to distribute to members in the form of additional benefits.

*1.1.3 Medicare Advantage Payment Controversy*

Medicare risk adjustment and thereby revenue is driven by diagnoses, which provides an incentive to plans to capture additional diagnoses. (See [3-7].h) The additional revenue received by the plan is required to be returned to beneficiaries in the form of additional benefits (referred to later as "ancillary benefits"). Thus, plans have the means to provide additional benefits and services to members. Which additional benefits are valued by members is the subject of this study.

*1.1.4 Prior Studies*

The literature on member choice in Medicare is dominated by two streams of research: how seniors choose between traditional Medicare and Medicare Advantage [8-10] and the impact and attractiveness of MA to underserved populations [11, 12]. In a 2015 paper, Jacobs and Buntin [13] investigated whether Medicare beneficiaries chose plans on the basis of premium or benefits, finding that lower-income beneficiaries are sensitive to cost-sharing while more affluent members choose plans with wider networks. Starting in 2020 MA plans are able to offer a wider range of benefits to members; there is little research on which plans offer these benefits and which benefits are valued and used by members [14-16]. There is also limited research on management decision-making with regard to benefit offerings. One survey-based study is that of Shields-Zeeman [17]. In part this may be because of limited reporting and data availability, as noted in a GAO report [18].

*1.2 Study objective*

Our objective is to identify those plan attributes that correlate with market share, as a proxy for benefits that members value. What makes a plan desirable to a prospective member? Is it cost, the range of benefits, the quality (Star ratings) or "brand"? Plans may choose to supplement their offerings from a wide range of plan features (premiums, deductibles, co-pays, limits etc.) and added benefits (dental, vision, transportation, meals). While we hypothesize that generous benefits and higher Star ratings attract more enrollment, it is important for public policy to understand whether plan revenue is dedicated to benefits that members value.

## 2 Data

MA enrollment data are published at the County/State/plan benefit package (PBP) level. Modeling the entire U.S. with 3,142 counties is beyond the scope of this study and we focus on a single county (Hudson) in New Jersey. In Hudson, seven different insurers offer a range of plans with high penetration among Medicare eligibles and a low market share concentration. We retrieved MA plan data (enrollment) from the CMS website.[1]

*2.1 Enrollment Data Set*

Our enrollment dataset contains information about the unique contracts in Hudson County between 2018 and 2023. MA contracts are identified by contract numbers beginning with 'H' and plan numbers less than 800. 800 series plan IDs (Employer Group Waiver Plans) were excluded from our analysis as benefit data is not publicly available for these plans. Later years have a greater number of contracts than the earlier ones, indicating overall MA plan growth. New contracts are added over time, and older contracts are canceled. The lack of continuity of contracts needs to be considered in modeling because it can possibly affect the relationship between benefits and enrollment.

Figure 1 shows the MA enrollment growth in Hudson County 2018-23. The compound annual growth rate is 10.5%. Table 2 shows the distribution of plans by market share 2018-23. Table 3 shows the trend in MA plan membership between HMO and PPO plans 2018-23, demonstrating an increasing preference on the part of members for open-network PPO plans rather than HMOs.

---

[1] Centers for Medicare & Medicaid Services. CMS.gov, `http://www.cms.gov` (last accessed on 07/25/2023)

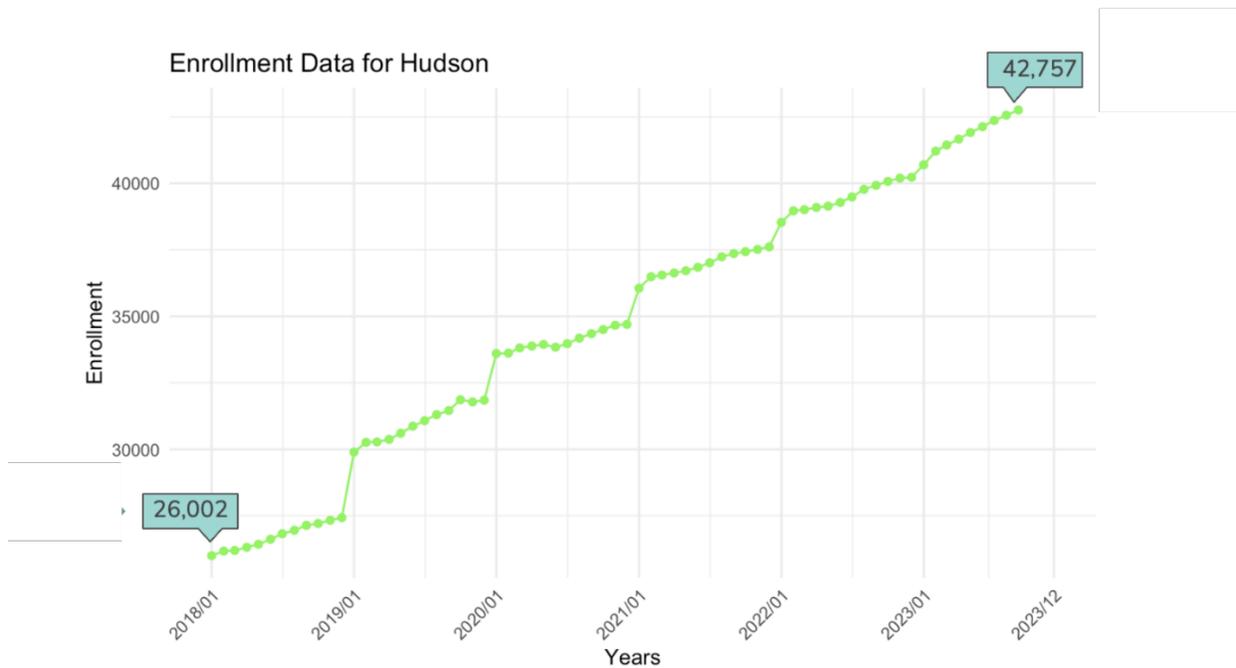

**Figure 1  MA Plan Enrollment  Hudson County 2018-2023[2]**

| Market share interval | Brand | Number of Plans | Market share interval | Brand | Number of Plans | Market share interval | Brand | Number of Plans |
|---|---|---|---|---|---|---|---|---|
| [0, 0.003) | Aetna | 19 | [0.003, 0.015) | Humana | 17 | [0.015, 0.27) | AARP | 20 |
| | BCBS | 8 | | Aetna | 15 | | Clover | 20 |
| | Humana | 7 | | Amerigroup | 9 | | BCBS | 9 |
| | AARP | 6 | | WellCare | 9 | | WellCare | 9 |
| | WellCare | 4 | | BCBS | 4 | | Aetna | 3 |
| | Clover | 2 | | Clover | 4 | | Amerigroup | 3 |
| | Amerigroup | 1 | | AARP | 2 | | Humana | 1 |
| | | 47 | | | 60 | | | 65 |

**Table 2  MA Plans by Marketshare 2018-23**

---

[2] Represents total membership in all MA plans, including DSNP, SNP and EGWP plans excluded from this study.

| | Plans | | Members | | | |
|---|---|---|---|---|---|---|
| Year | HMO | PPO | HMO | PPO | TOTAL | PPO % |
| 2018 | 4 | 11 | 9,147 | 8,130 | 17,277 | 52.9% |
| 2019 | 8 | 16 | 12,187 | 7,228 | 19,415 | 62.8% |
| 2020 | 8 | 18 | 9,853 | 11,212 | 21,065 | 46.8% |
| 2021 | 14 | 20 | 13,627 | 9,348 | 22,975 | 59.3% |
| 2022 | 16 | 19 | 17,156 | 6,999 | 24,155 | 71.0% |
| 2023 | 19 | 19 | 21,346 | 3,749 | 25,095 | 85.1% |
| | 69 | 103 | 83,316 | 46,666 | 129,982 | 64.1% |

**Table 3: Membership growth in HMO and PPO Plans 2018-23**

## 2.2 Benefits and Features Data

MA plans are required by law to offer all medically necessary services that Original Medicare covers. Plans may also offer some extra benefits that Original Medicare doesn't cover such as vision, hearing, and dental services. Plans are also required to use additional revenue for ancillary benefits such as meals, transportation and over-the-counter drug benefits. Plan features include monthly premiums, annual deductibles, coverage limits, cost-sharing etc. Detailed information about benefits and features was extracted at CMS.gov. An example of the dataset is provided in Table 4. We performed a preliminary review of the data to identify trends. Table 5 is a summary of some of the key values. Both Table 4 and Table 5 are included in the Appendix.

The majority of plans in all years offer Prescription Drugs. (Not all members purchase plans with drug benefits because Medicare Part D plans are available for separate purchase or from a former employer.) Member cost-sharing has been relatively stable over time, with some increase in Maximum out-of-pocket amount. Conversely PCP and specialist copays have declined over time, with minor increase in urgent care copays.

## 3 Data Preprocessing

There are no missing values in the dataset. We first removed some features for the following rationale: 1) lack of variability 2) strong multicollinearity 3) information redundancy. Specifically, features like "d12_in-Home_Safety_Assessment" and "d25_Telemonitoring_Services", which have zero variance were not considered as useful predictors and removed. We checked multicollinearity among all 65 features by visualizing the pairwise correlation matrix and computing variance inflation factors among all features. Features such as "Rehabilitation_services_physical" and "Copay_Plan" which are highly correlated with other features were removed to mitigate the impact of multicollinearity. Finally, features like "In(network)_Inpatient_coverage" and "Out (of network)_Inpatient_coverage" were aggregated into new normalized features to avoid duplicate information, and they were removed after we created corresponding new normalized features. The variable "Membership" was removed in favor of a new variable "market share." Finally, ID columns used to merge raw data files were removed.

Before modeling, to reduce the influence of outliers we applied min-max normalization to continuous features such that each continuous feature will be within the range 0 and 1. For categorical features, e.g. Star Rating, Year and Brand, one-hot encoding was used to represent categories as 0 or 1, and 1 was used whenever an observation was in that category.

**4 Model**

The final dataset contains 172 observations and 42 features. With such a small data set, using many features will very likely lead to an overfitting problem. More importantly, if we keep all features in the model, interpretability of each feature's coefficient will be influenced by collinearity. We use the Lasso (Least Absolute Shrinkage and Selection Operator) model [19] to shrink coefficients of features. If the coefficient of a feature is shrunk to zero, then the feature is considered as irrelevant and removed from the model. A shrinkage parameter is needed to adjust the magnitude of shrinkage of each coefficient. Five-fold cross validation is used to choose the optimal shrinkage parameter, which minimizes the cross-validation deviance. Once the shrinkage parameter is chosen, coefficients of some features shrink to zero. Our final model removes about half of the features and the interpretability of the results is improved.

Binomial logistic regression is a generalized linear regression model using a logistic function to estimate the occurring probability of the event of "success" in a binary response. Typically, the event of success is coded as 1; the event of failure is coded as 0. The model can be expressed as

$$P(Y = 1) = \frac{e^{\beta_0 + \beta_1 x_1 + \cdots + \beta_p x_p}}{1 + e^{\beta_0 + \beta_1 x_1 + \cdots + \beta_p x_p}}$$

Deviance is a goodness-of-fit statistic often used for logistic regression [20]. In a classification task Lasso adds an $L_1$ regularization or penalty term to the logistic regression model. $L_1$ regularization can force some coefficient estimates to be 0, and therefore it performs variable selection and leads to a more interpretable model. The coefficient estimates $\widehat{\beta_\lambda^L}$ minimize

$$\frac{1}{n} \text{Deviance}(\beta_0, \beta_1, \ldots, \beta_p) + \lambda \sum_{j=1}^{p} |\beta_j|.$$

In our study, we categorize the market share variable into 3 categories. The quantiles used to cut market share are 0, 0.003, 0.015 and 0.3. The numbers of plans for each of the 3 categories are 47, 60, 65, chosen so that the number of data points falling within each category is roughly balanced. For evaluation of the model, uniform distribution within categories avoids the bias that results when most data points are unbalanced.

We code the 3 categories: [0,0.003), [0.003, 0.015) and [0.015, 0.3) as 1, 2 and 3. For the multinomial outcome, the form of the probability in each category is extended from the binomial case and is expressed as:

$$P(Y = k) = \frac{e^{\beta_{0k} + \beta_{1k}x_1 + \cdots + \beta_{pk}x_p}}{\sum_{l=1}^{3} e^{\beta_{0l} + \beta_{1l}x_1 + \cdots + \beta_{pl}x_p}}$$

In this expression, $k \in \{1,2,3\}$. The model uses a partial Newton algorithm to estimate the coefficients and applies an $L_1$ penalty term to force some coefficients to be zero [21].

## 5 Results

Interpretation of the multinomial model is not straightforward. To simplify interpretation, we eliminate the intercept in each outcome category and interpret the influence of a predictor in terms of the ratio of probabilities. For example, to see how features affect the odds of a plan having a high market share vs. a low market share, we compute:

$$\frac{P(Y = 3)}{P(Y = 1)} = \frac{e^{\beta_{13} x_1 + \beta_{23}x_2 + \ldots + \beta_{p3}x_p}}{e^{\beta_{11} x_1 + \beta_{21}x_2 + \ldots + \beta_{p1}x_p}} = e^{(\beta_{13} - \beta_{11})x_1 + (\beta_{23} - \beta_{21})x_2 + \cdots + (\beta_{p3} - \beta_{p1})x_1}$$

That is, the ratio of two probabilities, (i.e., the odds) that a plan will be in one of the higher market share categories rather than the low market share category is determined by the (exponentiated) difference between coefficients of the covariates. To illustrate the application of the odds, consider the binary covariate HMO_PPO (HMO=1;PPO=0). The choice of HMO or PPO has no influence on Categories 1 or 2, but for Category 3, $e^{(\beta_{13} - \beta_{11})x_1 + (\beta_{23} - \beta_{21})x_2 + \cdots + (\beta_{p3} - \beta_{p1})x_1} = 0.863$, implying that an HMO plan is 0.863 times less likely to be in Category 3.

| Covariate | Coefficient 1 | Coefficient 2 | 2-1 | exp(2-1) | Coefficient 3 | 3-1 | exp(3-1) |
|---|---|---|---|---|---|---|---|
| **Brand** | | | | | | | |
| AARP | 0.000 | 0.000 | 0.000 | 1.000 | 1.257 | 1.257 | 3.515 |
| Aetna | 0.000 | 0.000 | 0.000 | 1.000 | -0.514 | -0.514 | 0.598 |
| Amerigroup | 0.000 | 0.528 | -0.528 | 0.590 | 1.000 | 1.000 | 1.000 |
| Clover | 0.000 | -0.316 | 0.316 | 1.371 | 0.184 | 0.184 | 1.203 |
| Humana | 0.000 | 0.000 | 0.000 | 1.000 | -0.546 | -0.546 | 0.579 |
| **Basic Benefits** | | | | | | | |
| HMO_PPO | 0.000 | 0.000 | 0.000 | 1.000 | -0.148 | -0.148 | 0.863 |
| Premium | 0.412 | 0.000 | -0.412 | 0.662 | -0.170 | -0.582 | 0.559 |
| Annual_Deductible | 0.000 | 0.000 | 0.000 | 1.000 | -0.736 | -0.736 | 0.479 |
| Inpatient days subject to deductible | 0.000 | 0.000 | 0.000 | 1.000 | 0.729 | 0.729 | 2.073 |
| In n/w_Inpatient_Unlimited Days | 0.303 | -0.020 | -0.323 | 0.724 | 0.000 | -0.303 | 0.739 |
| Out n/w_Inpatient_Unlimited Days | 0.000 | 0.000 | 0.000 | 1.000 | 0.250 | 0.250 | 1.284 |
| In n/w_Hearing_Exam copay | 0.000 | 0.000 | 0.000 | 1.000 | -1.229 | -1.229 | 0.293 |
| Emergency_Care copay per visit | 0.189 | -0.234 | -0.423 | 0.655 | 0.000 | -0.189 | 0.828 |
| Urgent_Care_Min copay per visit | 0.000 | -0.112 | -0.112 | 0.894 | 0.000 | 0.000 | 1.000 |
| Mental_services max. copay per visit | 0.000 | 0.000 | 0.000 | 1.000 | -0.273 | -0.273 | 0.761 |
| In n/w_Ambulance copay | 0.000 | 0.677 | 0.677 | 1.967 | 0.000 | 0.000 | 1.000 |
| In n/w_Doctor_visit_primary copay | 0.604 | -1.800 | -2.404 | 0.090 | 0.000 | -0.604 | 0.547 |
| Rehab_s_occupational copay | 0.020 | 0.000 | -0.020 | 0.981 | 0.000 | -0.020 | 0.981 |
| Drug_Coverage | -3.947 | 0.000 | 3.947 | 51.805 | 0.382 | 4.330 | 75.911 |
| Tier1 (Generic) copay | 0.000 | 0.000 | 0.000 | 1.000 | 0.031 | 0.031 | 1.032 |

| Covariate | Coefficient 1 | Coefficient 2 | 2-1 | exp(2-1) | Coefficient 3 | 3-1 | exp(3-1) |
|---|---|---|---|---|---|---|---|
| **Ancillary Benefits** | | | | | | | |
| d1_OTC_Drug_Benefits | 0.000 | 0.000 | 0.000 | 1.000 | 0.000 | 0.000 | 1.000 |
| d2_Meals_for_Short_Duration | 0.000 | 0.000 | 0.000 | 1.000 | -1.012 | -1.012 | 0.364 |
| d3_Annual_Physical_Exams | 0.401 | 0.000 | -0.401 | 0.670 | -0.652 | -1.053 | 0.349 |
| d4_Telehealth | 0.000 | 0.000 | 0.000 | 1.000 | 0.000 | 0.000 | 1.000 |
| d5_WW_Emergency_Transportation | 0.000 | 0.000 | 0.000 | 1.000 | 0.000 | 0.000 | 1.000 |
| d6_WW_Emergency_Coverage | 0.000 | 0.000 | 0.000 | 1.000 | 0.000 | 0.000 | 1.000 |
| d7_WW_Emergency_Urgent_Care | 0.000 | 0.000 | 0.000 | 1.000 | 0.000 | 0.000 | 1.000 |
| d8_Fitness_Benefit | 1.126 | 0.000 | -1.126 | 0.324 | 0.000 | -1.126 | 0.324 |
| d9_In-Home_Support_Services | 0.000 | 0.000 | 0.000 | 1.000 | 0.000 | 0.000 | 1.000 |
| d13_Personal Emergency Response Svcs. | 0.000 | 0.560 | 0.560 | 1.751 | 0.000 | 0.000 | 1.000 |
| d20_Nutritional/Dietary_Benefit | 0.374 | 0.000 | -0.374 | 0.688 | 0.000 | -0.374 | 0.688 |
| d26_Remote_Access_Technologies | 0.000 | -1.023 | -1.023 | 0.359 | 0.000 | 0.000 | 1.000 |
| d27_Counseling_Services | 0.000 | 0.000 | 0.000 | 1.000 | 0.000 | 0.000 | 1.000 |

Table 6: Results (Odds Ratios)

We computed the exponentiated difference between the coefficients of those features that remained from applying the multinomial Lasso model. Brand clearly has significant influence on plan choice: AARP and Clover plans are likely to be in the high market share category, while Aetna and Humana plans, and HMO plans are likely to be in lower market share categories. Notably it is the financial features that appear to have a significant influence on choice of plan, while ancillary benefits have little influence. The exception is drug coverage, the absence of which as one might expect, results in low market share because plans without drug coverage target a specific type of member.

Financial variables are influential. Exponentiated coefficient differences for the continuous financial covariates (premium, copay, deductible) are less than 1.00 implying that each dollar increase in these variables is likely to reduce the likelihood of a higher market share plan.

Few ancillary benefits influence market share: meals, fitness benefits, nutritional counseling and annual physical examinations are positively associated with market share. Technology (remote access and personal emergency response services) are modestly influential.

Some of the benefits that do not influence market share are surprising. Table 7 summarizes benefits that are not significant in our model. Benefits that are not influential include Star Ratings, which may be as a result of their effect being captured in the ancillary benefits that are funded by higher star ratings. Dental coverage and over-the-counter drug coverage are not influential, possibly because they are now offered by most plans. Other ancillary benefits may appeal to too small a subset of members to have significant influence on member plan choice.

| Benefits and Features that are not significant |
| --- |
| Star Ratings |
| In n/w Max. out of Pocket |
| Comprehensive dental coverage |
| OTC_Drug_Benefits |
| d10_Bathroom_Safety_Devices |
| d11_Health_Education |
| d12_In-Home_Safety_Assessment |
| d14_Medical_Nutrition_Therapy |
| d15_In-Home_Medication_Reconciliation |
| d16_Re-admission_Prevention |
| d17_Wigs_for_Hair_Loss |
| d18_Weight_Management |
| d19_Adult_Day_Health_Services |
| d21_Home-Based_Palliative_Care |
| d22_Support_for_Caregivers_of_Enrollees |
| d23_Smoking_Tobacco_Counseling |
| d24_Enhanced_Disease_Management |
| d25_Telemonitoring_Services |

Table 7: Features and Benefits that are not significant

## 6. Discussion

Medicare Advantage plans compete both on pricing and on the benefits that they offer. In some cases benefits are funded by additional revenue allocated based, among other things, on a plan's quality performance. Our interest is whether benefits that are not part of CMS's minimum are valued by members and drive members to choose plans that offer these benefits. We conclude that the ancillary benefits are relatively unimportant in influencing member choice. Instead, basic features such as premiums, copays, and deductibles appear to be more important.

## 7. Limitations and future work

Our study is based on a limited dataset covering plans between 2018-23 in a single county in New Jersey. Model accuracy could be different with a few changes in underlying data. More robust conclusions require analysis of a larger dataset. County or state itself could be an influential factor on plans' market share and future researchers can include these factors in the modeling. Interaction effects among important features can also be explored. However, for a small data set, including interaction effects adds model complexity and makes interpretation more difficult, and we have ignored interaction effects in this study.